\documentclass[%
 reprint,
groupedaddress,
showpacs,preprintnumbers,
amsmath,amssymb,
prl,
twocolumn
]{revtex4}
\usepackage{tipa}
\usepackage{pstricks}
\usepackage{graphics}
\usepackage{dcolumn}
\usepackage{bm}

\begin{document}
\preprint{Typeset using REVTEX4}
\title{Role of Alloying-Atom Size Factor and System Shape Factor in Energetics of bcc Fe under Macroscopic Deformation}
\author{Wei Liu,$^{1}$ Wei-Lu Wang,$^{1}$ Q. F. Fang,$^{1}$ C. S. Liu,$^{1,\ast}$
\footnotetext{*Author to whom correspondence should be addressed.
Email address: csliu@issp.ac.cn} Qun-Ying
Huang,$^{2}$ Yi-Can Wu$^{2}$} \affiliation{$^{1}$Key Laboratory of
Materials Physics, Institute of Solid State Physics, Chinese
Academy of Sciences, P. O. Box 1129, Hefei 230031, P. R. China \\
$^{2}$Institute of Plasma Physics, Chinese Academy of Sciences,
Hefei 230031, P. R. China}


\begin{abstract}
We present an \emph{ab initio} study of the effect of macroscopic
deformation on energetics of twelve alloying elements in bcc Fe
under three specially designed strain modes. We find that there
exists a universal linear relation of describing the volume
dependence of substitutional energy of alloying elements via
introducing two factors --- the system shape factor
($f_{\scriptsize{\mbox{ss}}}$) and the size factor of alloying
element $M$ ($\Omega^{M}_{\scriptsize{\mbox{sf}}}$):
$E_{\scriptsize{\mbox{sub}}} \sim
f_{\scriptsize{\mbox{ss}}}\Omega^{M}_{\scriptsize{\mbox{sf}}}V$.
$\Omega^{M}_{\scriptsize{\mbox{sf}}}$ well describes the effect of
intrinsic alloying-atom size and the influence of chemical
interaction with matrix atom, and $f_{\scriptsize{\mbox{ss}}}$
characterizes the degree of system lattice distortion under
deformation. This relation is further validated using the published
data of stained-modulated doping in GaP [Phys. Rev. Lett.
\textbf{105}, 195503(2010)].
\end{abstract}
\volumeyear{number}
\volumenumber{number}
\issuenumber{number}
\eid{identifier}
\date{\today}
\startpage{1}
\endpage{ }
\maketitle
Fe-based alloys are among the most widely used materials, of which
the ferritic martensitic steels represent a technologically
important class of materials with many applications in fission and
fusion energy systems. Alloying solutes in Fe-based alloys are of
great importance in modifying physical properties, and particularly
their contents strongly govern the mechanical performances such as
the resistance to hardening and embrittlement under neutron and
proton irradiation \cite{Leslie, Ishii, OlssonDomain}. It is found
that the optimal Cr content to improve one special property may be
different with that to improve another for bcc Fe based alloys,
e.g., the content of $2$ to $6\%$ Cr to reduce the irradiation
swelling whereas $9\%$ Cr to reduce ductile brittle transition
temperature (see Ref.\,3 and references therein). Most previous
investigations of alloying elements in ferritic martensitic steels
have been performed in macroscopically undeformed crystals
--- free of any macroscopic stain, except those induced by the
solutes or defects themselves. Nonetheless, as the structural
materials in fission and fusion reactors, the ferritic martensitic
steels experience both multiaxial loading and fairly high
temperatures, and the internal strain field becomes more complex due
to the presence of radiation damages. In fact, all materials usually
undergo macroscopic deformations due to externally applied loads or
in special in-service conditions. So in recent years many researches
have been made on the properties of solutes and point defects under
macroscopic deformations based on electronic structure calculations.
Gavini found that the volumetric strain associated with a
deformation largely governs the formation energies of monovacancies
and divacancies in Al and concluded the nucleation of these defects
is increasingly favorable under volumetric expansion \cite{Gavini}.
Chen \textit{et al}. reported that the strain has remarkable
influence on the stability, reorientation and migration of self
interstitial atoms (SIA) in bcc Fe \cite{Chen}: for instance,
uniaxial expansion induces a SIA spontaneous reorientation from
$<111>$ to $<100>$. Zhu \textit{et al}. found in zincblende GaP the
impurity formation energy changes monotonically in a linear fashion
with the applied external strain and thus proposed that the
intriguing unbounded strain-induced change in impurity formation
energy can be used effectively to enhance dopant solubility in a
wide range of semiconductors \cite{Zhu}. These results raise an
interesting question on how macroscopic deformation affects the
stability of alloying elements in Fe-based alloys, which to our
knowledge remains open. If the macroscopic deformation does
influence on the energetics of alloying elements, the aggregation
and precipitation may take place consequently, and thus the
mechanical performance may be degraded.

In this Letter, we present an \emph{ab initio} study of the effect
of macroscopic deformation on substitutional energy of twelve
alloying elements (Al, Co, Cr, Cu, Mo, Nb, Ni, Si, Ta, Ti, V, and W
\cite{supplementary}) in bcc Fe under three different strain modes.
Hydrostatic strain mode is to explore the pure volume change effect,
and a set of submodes in both normal strain and shear stain modes
are to explore the lattice distortion effect. We find that the
macroscopic deformation effect on substitutional energy of alloying
elements is governed linearly and universally by two factors, the
system shape (lattice distortion) factor and the size factor of
alloying element. The size factor of alloying element well describes
the effect of intrinsic alloying-atom size and the influence of
chemical interaction with matrix atom, and the system shape factor
characterizes the degree of lattice distortion under deformation.

The present calculations are performed within spin-polarized density
functional theory as implemented in the Vienna \emph{ab initio}
simulation package (VASP) \cite{Kresse}. The interaction between
ions and electrons is described by the projector augmented wave
(PAW) method \cite{KresseBlohl}. Exchange and correlation functions
are taken in a form proposed by Perdew and Wang (PW91)
\cite{PerdewWang} within the generalized gradient approximation
(GGA). The supercell approach with periodic boundary conditions is
used. The supercell contains $127$ Fe atoms and one alloying atom.
We used an energy cutoff of 350 eV for the plane-wave expansion of
the wave functions and a $3\times 3 \times 3$ \textit{k}-point mesh
for Brillouin zone sampling according to the Monkhorst-Pack scheme.
All atomic relaxation calculations are performed at constant volume
and shape using the conjugate gradient algorithm.
For simplicity the side length of the forementioned
bcc cubic supercell in equilibrium is set to be four times of the
optimized lattice constant of bcc Fe ($2.833$ \r{A}). In the
calculation of substitutional energy the following formula is used:
\begin{equation}
\small{E_{\scriptsize{\mbox{sub}}}} = \small{E(n\mbox{Fe}+1M)} -
\small{\frac{n}{n+1}}\times
\small{E((n+1)\mbox{Fe})}-\small{E(M)},\label{eq1}
\end{equation}
where $E(n\mbox{Fe}+1M)$ is the total energy of a bcc cubic
supercell containing $n$ Fe atoms and one alloying atom $M$,
$E((n+1)Fe)$ is the total energy of the same bcc cubic supercell
filled with only Fe atoms under the same deformation, and $E(M)$ is
the energy per atom of pure crystal $M$ in its stablest phase.

\begin{table}
\caption{\label{TABLE I}Strain components (\%) of each submode in NS and SS.}
\begin{ruledtabular}
\begin{tabular}{lccc}
&\multicolumn{3}{c}{$e_{11}$,~$e_{22}$,~$e_{33}$~(NS1--4)/$e_{11}$,~$e_{33}$,~$e_{12}$~(SS89--85)}\\\cline{2-4}
\raisebox{5pt}[0pt]&\raisebox{-1pt}[0pt]{Vol.
contracted}&\raisebox{-1pt}[0pt]{Vol.
unchanged}&\raisebox{-1pt}[0pt]{Vol. expanded}
\\&by $3.0\%$& &by $3.0\%$\\
\hline
NS1 & $-1.98$, $-1.0$, $-0.01$&$-0.99$, $0$, $1.0$& $0$,     $1.0$, $2.01$\\
NS2 & $-2.94$, $-1.0$, $0.98$& $-1.96$, $0$, $2.0$& $-0.98$, $1.0$, $3.02$\\
NS3 & $-3.88$, $-1.0$, $1.97$& $-2.91$, $0$, $3.0$& $-1.94$, $1.0$, $4.03$\\
NS4 & $-4.81$, $-1.0$, $2.96$& $-3.85$, $0$, $4.0$& $-2.88$, $1.0$, $5.04$\\
SS89 & $-0.996$, $-1.0$, $0.864$&$0.004$,$0$, $0.873$&$1.004$,$1.0$, $0.881$\\
SS88 & $-0.985$, $-1.0$, $1.73$& $0.015$, $0$, $1.75$& $1.02$, $1.0$, $1.76$\\
SS87 & $-0.966$, $-1.0$, $2.59$& $0.034$, $0$, $2.62$& $1.03$, $1.0$, $2.65$\\
SS85 & $-0.906$, $-1.0$, $4.33$& $0.096$, $0$, $4.37$& $1.10$, $1.0$, $4.41$\\
\end{tabular}
\end{ruledtabular}
\end{table}

In this work, the macroscopic deformation is described by a
macroscopic strain tensor with the coordinate axis 1, 2 and 3 being
chosen along [100], [010] and [001] directions of the perfect bcc
crystal, respectively. Because of strain tensor belonging to a
six-dimensional space, a complete characterization of deformation
effect on the energetics of alloying atoms is beyond reach. Here we
focus on three different modes of strain: hydrostatic strain (HS),
normal strain (NS) and shear strain (SS) within relative volume
variation from $-3.0\%$ to $3.0\%$. Both NS and SS modes include
four submodes: NS1--4 and SS89--85, as listed in Table~\ref{TABLE
I}. Here, for example, SS89 denotes that the angle between
two adjoining surfaces is changed to $89^{\circ}$ from initial value
$90^{\circ}$. Note that, 1) in NS/SS mode the cube symmetry of the
system changes into orthogonality/monoclinism; 2) in each submode of
NS and SS a set of strains are applied in such a way that leads to
different volumetric change but the system being kept
\textit{similar shape}, in other words, the ratio of three edges and
the angle between any two intersected edges of the
supercell are unchanged. For example, in submode NS1 the system is
in rectangular parallelepiped shape and the ratio of three different
edges is fixed at $0.99:1.0:1.01$. From NS4 to NS1 (from SS85 to
SS89) the system shape approaches the perfect cube step by step.
For all submodes of SS mode the following
equations exist: \begin{math}e_{11}=e_{22}\end{math},
\begin{math}e_{23}=e_{31}=0\end{math}. Of these submodes the strain
components with relative volume changes of $-3.0\%$(contracted),
$0\%$(kept at equilibrium) and $3.0\%$(expanded) are summarized in
Table~\ref{TABLE I}.

\begin{figure}[htb]
\includegraphics{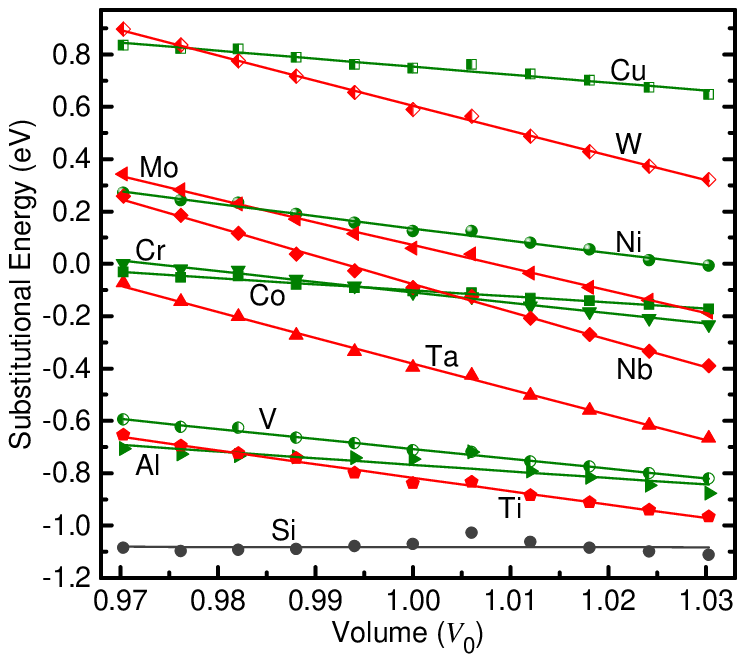}
\caption{\label{FIG. 1} (Color online) Substitutional energy versus
volume for twelve different alloying elements in HS mode.
Symbols represent the calculated results and solid lines are
the corresponding linear fits.}
\end{figure}

\begin{figure}[htb]
\includegraphics{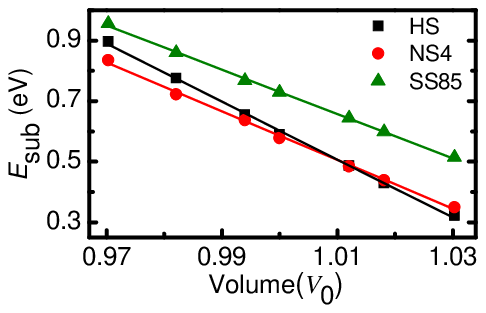}
\caption{\label{FIG. 2} (Color online) Substitutional energy
$E_{\scriptsize{\mbox{sub}}}$ of W versus volume in HS, NS4 and
SS85. Lines are linear fits to the data.}
\end{figure}

We first investigate the isotropic volumetric deformation influence
on substitutional energy of alloying elements in HS mode.
Figure\,\ref{FIG. 1} displays the substitutional energy as a
function of volume ranging from $0.97V_{0}$ to $1.03V_{0}$ in HS
mode. In our present work $V_{0}$ is the equilibrium volume of the
supercell with 128 Fe atom. As can be seen, the substitutional
energy of Si is insensitive to isotropic deformation, while with the
increase of volume those of Al, Co, Cr, Cu, Ni, and V decrease
slowly and those of Mo, Nb, Ta, Ti and W decrease strongly. The
results suggest that in HS mode the stability of the alloying
elements studied except Si are considerably influenced by
hydrostatic volumetric deformation, and these alloying atoms except
Si prefer strongly to stay in the expansion zone, which in turn
changes the distributions of alloying atoms in bcc Fe, considering
the direct correlation between substitutional energy and solubility.
For every studied alloying element the substitutional energy depends
linearly on the volume with a negative slope (except Si with a much
small positive slope), which is dependent on alloying element.
Interestingly, when examining the calculated substitutional energies
of each alloying element in each submode of NS/SS with volumes of
$0.97V_{0}$, $V_{0}$, and $1.03V_{0}$, we find the similar linear
volume dependence as in HS mode but with an increased slope. The
obtained linearity has been further checked via the added
calculations for Al, Cr, and W in NS1, NS4, SS89, and SS85 submodes
with volume of $0.982V_{0}$, $0.994V_{0}$ and $1.018V_{0}$, and here
only the results of W in NS4 and SS85 submodes are presented in
FIG.\,\ref{FIG. 2}. Figure\,\ref{FIG. 2} clearly indicates that,
there does exist a similar linear volume dependence of
substitutional energy in submodes in NS and SS modes as in HS mode
but with a larger slope. For example, in HS the isotropic volumetric
increase from $0.97V_{0}$ to $1.03V_{0}$ leads to a substitutional
energy decrease of 0.234\,eV for Cr and 0.575\,eV for W, while in
NS4/SS85 the substitutional energy decrease are 0.162/0.116\,eV for
Cr and 0.486/0.442\,eV for W. Note that ${\Delta}E_{sub}$ ($=
E_{\scriptsize{\mbox{sub}}}|_{V=1.03V_{0}} -
E_{\scriptsize{\mbox{sub}}}|_{V=0.97V_{0}}$) divided by $0.06V_{0}$
is approximately identical with the element-dependent slope of the
forementioned linearity between the substitution energy and volume.
These results tell us that in any submode of NS/SS the volumetric
deformation effect on the stability of the studied alloying elements
except Si is noticeably reduced compared with that in HS mode. Our
observed linear volume dependence of substitutional energy are in
good agreement with the reported linear behavior of impurity
formation energy in semiconductors under hydrostatic strain by Zhu
\textit{et al} \cite{Zhu}. They found that the change of impurity
formation energy is a linear function of volume and the slope is
proportional to the effective size difference between the host and
the dopant atom; and trends of the strain-induced change of impurity
formation energy for the biaxial and the hydrostatic strain are
about the same, except that the effect of the biaxial strain is
smaller. Thus, the deformation in NS and SS modes can modulate to
some extent the volume dependence of substitution energy compared to
pure isotropic volumetric change in HS mode.

\begin{figure}[htb]
\includegraphics{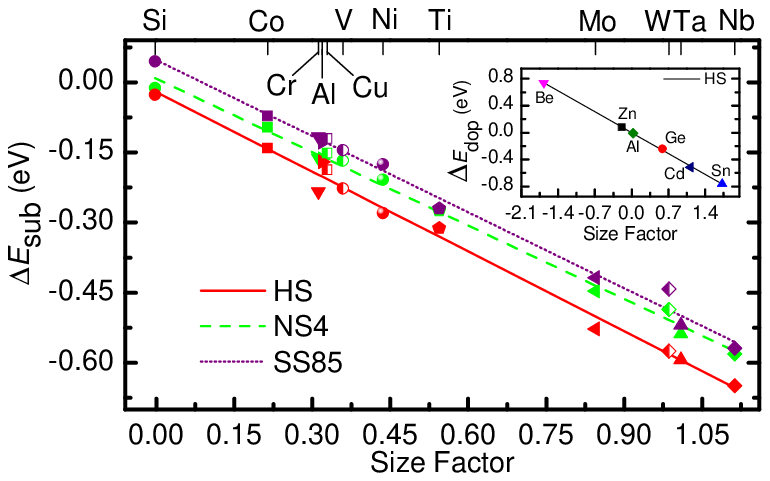}
\caption{\label{FIG. 3} (Color online) Substitutional energy change
${\Delta}E_{\mbox{\scriptsize{sub}}}$ versus size factor in HS, NS4
and SS85 with volume change from $0.97V_{0}$ to $1.03V_{0}$ for
twelve alloying elements. Inset: The doping energy difference
resulted from the strain changes between $-2\%$ and $2\%$ in HS mode
in GaP versus size factor of dopant. Lines are linear fits to the
data.}
\end{figure}

Above results reveal for every studied alloying element the
existence of linear volume dependence of the substitutional energy
in the same deformation mode (i.e.,the system is kept \emph{a
similar shape} such as in the HS mode and in the submode of NS/SS
mode), but the slope is dependent on the alloying element. As can be
seen from FIG.\,\ref{FIG. 1}, for the alloying atom with much larger
radius the slope is smaller than that with a smaller atomic radius
(e.g., Ta and Si). This implies that the difference in atomic size
between the alloying atom and the host is a key factor for further
exploring the linear volume dependence of the substitutional energy.
Recently, Olsson \textit{et al}. proposed a special method to
calculate the size factor ($\Omega^{M}_{\scriptsize{\mbox{sf}}}$) of
single solute atom $M$ in bcc Fe matrix, and found that the
calculated $\Omega^{M}_{\scriptsize{\mbox{sf}}}$ can reflect not
only the effect of the solute atom size but also the influence of
the chemical interaction with the matrix \cite{Olsson}. Following
this method, we first obtain the bulk modulus $B$ for every binary
alloy by fitting the linear relation between pressure and volume
(${\Delta}P=-B{\Delta}V/V_{0}$) in HS mode with volume varying from
$0.97V_{0}$ to $1.03V_{0}$. Then the size factor of the alloying
atom is calculated by
\begin{equation}
\Omega^{M}_{\scriptsize{\mbox{sf}}}=
-V_{0}{\Delta}P_{\scriptsize{\mbox{ex}}}/(\Omega_{\scriptsize{\mbox{Fe}}}B),\label{eq2}
\end{equation}
where ${\Delta}P_{\scriptsize{\mbox{ex}}}$ is the excess pressure of
the binary alloy and $\Omega_{\mbox{\scriptsize{Fe}}}$ is the atomic
volume of bcc Fe. ${\Delta}P_{\scriptsize{\mbox{ex}}}$ is calculated
by subtracting the external pressure of pure bcc Fe matrix in HS
from that of the binary alloy in HS with the same equilibrium volume
$V_{0}$. If the $\Omega^{M}_{\scriptsize{\mbox{sf}}}$ of alloying
atom correctly describe the intrinsic size difference and the change
in atomic interaction, then there may exists a general (or
universal) law of the macroscopic deformation effect on substitution
energy. So we replot the substitution energy changes
${\Delta}E_{\scriptsize{\mbox{sub}}}$ of twelve alloying elements as
a function of $\Omega^{M}_{\scriptsize{\mbox{sf}}}$ in
FIG.\,\ref{FIG. 3}. To our surprise, as displayed in FIG.\,\ref{FIG.
3} it is found that in those modes/submodes whose system being kept
a similar shape ${\Delta}E_{\scriptsize{\mbox{sub}}}$ exhibits a
universal dependence on the $\Omega^{M}_{\scriptsize{\mbox{sf}}}$:
for all studied alloying elements
${\Delta}E_{\scriptsize{\mbox{sub}}}$ (i.e. the slope of linear
volume dependence of $E_{\scriptsize{\mbox{sub}}}$) is approximately
inverse proportional to $\Omega^{M}_{\scriptsize{\mbox{sf}}}$. In
Ref.\,6, it is obtained as a first order approximation that
\begin{math}E_{\mbox{\scriptsize{dop}}}{\sim}\alpha(V_{\mbox{\scriptsize{host}}}-V_{\mbox{\scriptsize{host}}+\mbox{\scriptsize{dopant}}})V\end{math},
where $\alpha$ is the elastic constant,
$V_{\scriptsize{\mbox{host+dopant}}}$ is the equilibrium volume of
the system with the dopant and $V_{\scriptsize{\mbox{host}}}$ is the
equilibrium volume of the host lattice. This formula can also well
describe the linear relation of substitutional energy versus volume
in FIG.\,\ref{FIG. 1}, and the slope is determined by
$\alpha(V_{\scriptsize{\mbox{host}}}-V_{\scriptsize{\mbox{host+M}}}$).
Based on the above discussion
$\alpha(V_{\scriptsize{\mbox{host}}}-V_{\scriptsize{\mbox{host+dopant}}})$
in Ref. 6 corresponds to $\Omega^{M}_{\scriptsize{\mbox{sf}}}$ in
present work. Therefore, using Eq.\,\ref{eq2} (where
${\Delta}P_{\scriptsize{\mbox{ex}}}$ is replaced by
$-B{\Delta}V_{\scriptsize{\mbox{ex}}}/V_{0}$ and
$\Omega_{\scriptsize{\mbox{Fe}}}$ is replaced by
$\Omega_{\scriptsize{\mbox{Ga}}}$) and the data summarized in Table
I in Ref. 6 on strain-enhanced doping in semiconductors, we
calculate $\Omega^{D}_{\scriptsize{\mbox{sf}}}$ of doping elements
in GaP. The doping energy difference
${\Delta}E_{\scriptsize{\mbox{dop}}}$ resulted from the strain
change between -2\% and 2\% in HS mode as reported in FIG. 1 in Ref.
6 is redisplayed as a function of the calculated
$\Omega^{D}_{\scriptsize{\mbox{sf}}}$ in the inset of FIG. 3. This
inset figure clearly indicates that for doping of zincblende GaP the
doping energy changes resulted from hydrostatic strain show a
similar universal dependence on the
$\Omega^{D}_{\scriptsize{\mbox{sf}}}$, though the doping energies of
these elements exhibit different, even opposite volume dependence.
Thus the $\Omega^{M/D}_{\scriptsize{\mbox{sf}}}$ is a suitable
factor which well describes the intrinsic size difference and the
change in atomic interaction between alloying/doping atom with host
atom and produces a universal effect of pure volumetric deformation.
In addition, as forementioned and shown in FIG. 2 the
element-dependent slope of the linearity between the substitution
energy and volume is larger in any submode of NS/SS than in HS mode,
suggesting that the system shape factor
($f_{\scriptsize{\mbox{ss}}}$) should be introduced to describe the
lattice distortion of system under strained conditions. So the
substitutional energy change due to macroscopic deformation can be
described universally by the following formula:
\begin{equation}
E_{\scriptsize{\mbox{sub}}}=f_{\scriptsize{\mbox{ss}}}\Omega^{M}_{\scriptsize{\mbox{sf}}}V+I.\label{eq3}
\end{equation}
Here, the constant/intercept $I$ is dependent on both the alloying
element and the strain submode. By linear fitting of the calculated
${\Delta}E_{\scriptsize{\mbox{sub}}}$ versus
$\Omega^{M}_{\scriptsize{\mbox{sf}}}$ the
$f_{\scriptsize{\mbox{ss}}}$ of each NS and SS submode is obtained
and listed in Table~\ref{TABLE II}. As can be seen, from NS4 to NS1
and from SS85 to SS89 the system lattice distortion is weakened, the
system shape factor values become smaller ant approaches the value
in HS mode where no lattice distortion. This result suggests that
the system shape factor can well describe the system lattice
distortion. Therefore, the macroscopic deformation effect on
substitutional energy can be divided into two parts: the system
shape distortion effect described by the
$f_{\scriptsize{\mbox{ss}}}$ factor and the pure volumetric change
effect generalized by the $\Omega^{M}_{\scriptsize{\mbox{sf}}}$ for
different alloying elements. In shape/lattice distorted systems the
pure volumetric change effects on substitutional energy is enlarged
for the alloying element who has negative size factor, and is
reduced for the one who has positive size factor compared with the
cases in perfect cubic systems.

\begin{table}
\caption{\label{TABLE II}The system shape factor of each NS/SS
submode in unit of eV$\mbox{\r{A}}^{\scriptsize{-3}}$. The system
shape factor in HS mode is -2.51
eV$\mbox{\r{A}}^{\scriptsize{-3}}$.}
\begin{ruledtabular}
\begin{tabular}{ccccccccc}
Mode&NS4&NS3&NS2&NS1&SS85&SS87&SS88&SS89\\
\hline
$f_{\scriptsize{\mbox{ss}}}$&\small{$-2.31$}&\small{$-2.38$}&\small{$-2.40$}&\small{$-2.48$}&\small{$-2.39$}&\small{$-2.40$}&\small{$-2.45$}&\small{$-2.48$}\\
\end{tabular}
\end{ruledtabular}
\end{table}

In summary, the \emph{ab initio} calculations reveal that within
volume change ranging from $0.97V_{0}$ to $1.03V_{0}$ the
macroscopic deformation effect on the substitutional energy of
alloying elements can be divided into pure volume change effect and
lattice distortion effect and be well described by a universal and
simple relation: $E_{\scriptsize{\mbox{sub}}}\sim
f_{\scriptsize{\mbox{ss}}}\Omega^{M}_{\scriptsize{\mbox{sf}}}V$.
This relation should be beneficial to understanding the mechanical
performance degrading of Fe-based alloys under deformation. We
believe that this relation can be generalized to other systems in
studying substitutional or doping energy change under deformed
condition. In addition, in the lattice distorted systems the pure
volume change effects on substitutional energy is enhanced for those
alloying elements who have negative size factor, and is reduced for
those who have positive size factor compared to the cases without
lattice distortion, which may be instructive to design better
materials.

\begin{acknowledgments}
This work was supported by the Innovation Program of Chinese Academy
of Sciences (Grant Nos.: KJCX2-YW-N35 and and XDA03010303) and the
National Magnetic Confinement Fusion Program (Grant No.:
2009GB106005), and by the Center for Computation Science, Hefei
Institutes of Physical Sciences.
\end{acknowledgments}


\begin{thebibliography}{1}
\bibitem {supplementary}In hydrostatic strain
we calculated the substitutional energies of Mn: the value is almost
constant ($0.118\pm0.032$ eV) in the volume range from $0.970V_{0}$
to $1.024V_{0}$, but it changes to $-0.048$ eV at $1.03V_{0}$. We do
not know the whys of this unusual result of Mn. So the result of Mn
is not presented in this paper.
\end{thebibliography}
\end{document}